 \documentclass[preprint]{aastex631}

\graphicspath{{./}{figures/}}

\shorttitle{FAST search for circumstelar \ion{H}{1}: PNe at $|b|\geq10\arcdeg$}
\shortauthors{Ouyang et al.}

\begin{document}

\title{FAST search for circumstellar atomic hydrogen. III. 
Planetary nebulae at Galactic latitudes $|b|\geq10\arcdeg$}

\correspondingauthor{Yong Zhang}
\email{zhangyong5@mail.sysu.edu.cn}

\author[0000-0002-2762-6519]{Xu-Jia Ouyang}
\affiliation{School of Physics and Astronomy, Sun Yat-sen University, 2 Daxue Road, Tangjia, Zhuhai, Guangdong Province,  China}

\author[0000-0002-1086-7922]{Yong Zhang}
\affiliation{School of Physics and Astronomy, Sun Yat-sen University, 2 Daxue Road, Tangjia, Zhuhai, Guangdong Province,  China}
\affiliation{Xinjiang Astronomical Observatory, Chinese Academy of Sciences, 150 Science 1-Street, Urumqi, Xinjiang 830011, People's Republic of China}
\affiliation{CSST Science Center for the Guangdong-Hongkong-Macau Greater Bay Area, Sun Yat-Sen University, Guangdong Province, China}
\affiliation{Laboratory for Space Research, The University of Hong Kong, Hong Kong, China}

\author[0000-0002-3171-5469]{Albert Zijlstra}
\affiliation{Department of Physics and Astronomy, The University of Manchester, Manchester M13 9PL, UK}
\affiliation{Laboratory for Space Research, The University of Hong Kong, Hong Kong, China}

\author[0000-0002-4428-3183]{Chuan-Peng Zhang}
\affiliation{National Astronomical Observatories, Chinese Academy of Sciences, Beijing 100101, China}
\affiliation{CAS Key Laboratory of FAST, National Astronomical Observatories, Chinese Academy of Sciences, Beijing 100101, China}

\author[0000-0003-3324-9462]{Jun-ichi Nakashima}
\affiliation{School of Physics and Astronomy, Sun Yat-sen University, 2 Daxue Road, Tangjia, Zhuhai, Guangdong Province,  China}
\affiliation{CSST Science Center for the Guangdong-Hongkong-Macau Greater Bay Area, Sun Yat-Sen University, Guangdong Province, China}
                
\author[0000-0002-2062-0173]{Quentin A Parker}
\affiliation{Faculty of Science, The University of Hong Kong, Chong Yuet Ming Building Pokfulam Road, Hong Kong, China}
\affiliation{Laboratory for Space Research, The University of Hong Kong, Hong Kong, China}
\affiliation{CSST Science Center for the Guangdong-Hongkong-Macau Greater Bay Area, Sun Yat-Sen University, Guangdong Province, China}

\author[0000-0003-2090-5416]{Xiao-Hu Li}
\affiliation{Xinjiang Astronomical Observatory, Chinese Academy of Sciences, 150 Science 1-Street, Urumqi, Xinjiang 830011, People's Republic of China}



\begin{abstract}

 The detection of circumstellar atomic hydrogen (\ion{H}{1})  via the 21\,cm line remains a persistent challenge in planetary nebula (PN) studies, primarily due to contamination from ubiquitous interstellar \ion{H}{1} 21\,cm emission. In this paper, we report the results of a \ion{H}{1} survey of 12 high surface brightness PNe located at  Galactic latitudes $|b|\geq10\arcdeg$, performed with the Five-hundred-meter Aperture Spherical radio Telescope (FAST), which is currently the most sensitive telescope in 
the $L$ band. Although the contamination from interstellar
emission is still severe, we detect or tentatively 
detect circumstellar \ion{H}{1} 21\,cm absorption associated with  
 two PNe: NGC\,6905 and NGC\,7662.
\ion{H}{1} exhibits a comparatively high detection frequency in bipolar PNe.
We estimate the
 \ion{H}{1} masses of the two PNe to range from
 0.01 to 0.14 $M_{\sun}$, resulting in  
atomic-to-ionized hydrogen ratios of 0.02--0.3.
The \ion{H}{1} shells have dynamical ages of
2100--2400 years. Our measurements confirm 
previous findings that the optical depth of 
\ion{H}{1} decreases with increasing linear radius
of the nebula. 
The mass loss rates traced by the \ion{H}{1} absorption
are larger than  10$^{-5}$\,M$_\sun$ yr$^{-1}$, indicating that they originate from the superwind phase at the tip of the asymptotic giant branch.

\end{abstract}


\section{Introduction} \label{sec:intro}

When low- and intermediate-mass stars (0.8 to 8 $\rm M_{\sun}$) 
evolve through the asymptotic giant branch (AGB) and post-AGB
phases, they eject much of their mass into the interstellar medium (ISM).
 Even after a significant portion of this mass has dissipated, there is a final expulsion of an envelope composed of circumstellar materials. This envelope is subsequently ionized by the central star, which has been heating up as it evolves towards becoming a white dwarf. At this particular stage, it gives rise to what is known as an optically visible planetary nebula (PN).
The study of PNe greatly aids our understanding not 
just of late-stage stellar evolution per se but of the stellar mass feedback mechanism for
galaxy evolution for the dominant low mass stellar population \citep[see][for an excellent recent review]
{2022PASP..134b2001K}.
 A persistent challenge in PN research is the significant mass discrepancy between observed nebular systems and theoretical predictions. The total mass of ionized nebulae and their central stars \citep[typically $\le$ 1.5 $\rm M_{\sun}$,][]{2012A&A...541A.112K} falls substantially below the theoretical upper mass limit for progenitor stars. This inconsistency persists even when accounting for the faint yet extended AGB halos observed in some PNe \citep[the so-called ``PN missing mass problem",][]{1994PASP..106..344K}.
Extensive, deep, observations show that the combined inventory of surrounding
molecules, dust, and ionized gas is insufficient to 
account for the mass lost by the central star when finally emerging as a PN 
\citep{1996A&A...315..284H,2007MNRAS.381..117P,2012A&A...541A.112K,2020MNRAS.491..758A,2022NatAs...6.1421D,2022ApJ...940...54S,2023MNRAS.526.5386D} with the ionized PN shell itself typically being only $\sim0.1M_{\sun}$.
Locating the missing mass has been problematic. 
 Detecting 21\,cm atomic hydrogen (\ion{H}{1}) 
line\footnote{Throughout this paper,
as a simplification, we use ``\ion{H}{1} line''
to refer to the atomic hydrogen line at 21\,cm.} from the spin-flip transition
remains exceptionally challenging in PNe.
This is largely because of the considerable contamination by the omnipresent
21\,cm emission of the ISM along the lines of sight in the absence of very high resolution kinematic mapping.

Investigating the complete mass loss history of a star poses significant challenges. 
Historically, researchers have primarily relied on CO emission in the submillimeter wavebands or dust scattering 
within the circumstellar envelope (CSE) to study stellar mass loss. 
However, these methods come with inherent limitations.
For instance, CO, even in evolved stars with high mass loss rates ($\sim10^{-5}$ $\rm M_{\sun}\,yr^{-1}$), traces 
scales of only $\sim10^4$ AU \citep[e.g.,][]{2019A&A...625A..81S,2020A&A...640A.133R}. 
Similarly, dust-scattered light covers spatial scales comparable to those traced by CO \citep[e.g.,][]
{2006A&A...452..257M,2013A&A...551A.110M}.
Consequently, tracers such as CO and dust do not reveal the more extended CSEs surrounding evolved stars.
Interestingly, far-ultraviolet (FUV) observations have uncovered more extended structures around some AGB stars, due 
to interactions between molecular winds and the ISM \citep[e.g.,][]
{2007Natur.448..780M,2023AJ....165..229S}.
These structures are an order of magnitude larger than those traced by CO and dust 
\citep{2007Natur.448..780M,2010ApJ...711L..53S,2014AJ....148...74S,2023MNRAS.522..811O,2023AJ....165..229S,2024MNRAS.527.4730G}. 
This implies that stellar mass loss extends back much further in time than can be traced by CO and dust.
The luminosity observed in the FUV is thought to arise from the collisional excitation of H$_2$ molecules by hot 
($\sim$30 eV) electrons \citep{2007Natur.448..780M,2010ApJ...711L..53S,2014AJ....148...74S,2023AJ....165..229S}. 
According to this hypothesis, the dissociation of H$_2$ likely results in the production of \ion{H}{1} as a 
byproduct. \citet{1983MNRAS.203..517G} suggest that for a photospheric temperature larger than 2500 K, the hydrogen 
flowing from an AGB star is mainly atomic. This has been confirmed by the detection of  \ion{H}{1} emission
associated with AGB envelopes \citep{2006AJ....132.2566G,2013AJ....145...97M,2024A&A...692A..54G}. 
Furthermore, during  PN evolution, \ion{H}{1} can be formed through photodissociation of H$_2$.
Therefore, \ion{H}{1} observations of PNe can provide significant information on their early mass-loss history. 
To the best of our knowledge, only five PNe, specifically IC\,418, NGC\,7293, BD+30\arcdeg3639, KjPn\,8, and NGC\,6369, have been detected to date exhibiting \ion{H}{1} 21\,cm emission features \citep{1989ApJ...340..932T,1990ApJ...351..515T,2000RMxAA..36...51R,2002ApJ...574..179R,2006AJ....132.2566G}. Additionally, \ion{H}{1} absorption has been observed in ten PNe \citep{1982Natur.299..323R,1986ApJ...305L..85A,1987ApJ...314..572S,1987A&A...176L...5T,1990ApJ...351..515T,1995MNRAS.273..801G}.

The number of PNe exhibiting circumstellar \ion{H}{1} detections is exceedingly limited compared with the 
overall number of True (T), Likely (L) and Possible (P) Galactic PNe recorded in the Hong 
Kong/AAO/Strasbourg/H$\alpha$ PN database  \citep[HASH PN database,][]{2016JPhCS.728c2008P} which currently stands 
 at 3953 as at April 2025.
This precludes meaningful statistical study on the
relationship between the \ion{H}{1} content and the evolution and characteristics of the associated PN. 
 To this end, our research group has 
undertaken efforts to investigate \ion{H}{1} in PNe utilizing the Five-hundred-meter Aperture Spherical 
radio Telescope \citep[FAST,][]{2011IJMPD..20..989N}, which
provides the most sensitive $L$-band spectra of PNe to date. In Paper I \citep{2022ApJ...933....4O},
we reported the detection of a \ion{H}{1} absorption toward  IC\,4997, which originates from a shell having a 
dynamic age of $\sim990$ yr. In Paper II  \citep{2023ApJ...952..166O}, we reported a detection of extended \ion{H}
{1} emission surrounding the young PN BD+30\arcdeg3639, which may
arise from the ISM progressively swept out by the stellar wind. These results indicate that young PNe probably
contain a large amount of detectable  \ion{H}{1}.

Utilizing the detection of \ion{H}{1} absorption and emission
in IC\,4997 and BD+30\arcdeg3639 as a reference, we have performed a search of \ion{H}{1} in a sample  of high 
surface brightness, mostly compact PNe.
It can be considered as a development
of the pioneer work of \citet{1987ApJ...314..572S} at Arecibo. The more sensitive FAST observations can provide a 
tighter constraint on the \ion{H}{1} mass. Section 2 describes the observations and data
reduction. The results are reported in Section 3, where we estimate the masses or mass upper limits of \ion{H}{1}.
The implications of our observations on the PN evolution are discussed in Section 4, followed by the conclusions in 
Section 5.

\section{Observation and Data reduction} \label{sec:obse}

\begin{figure}[ht!]
\plotone{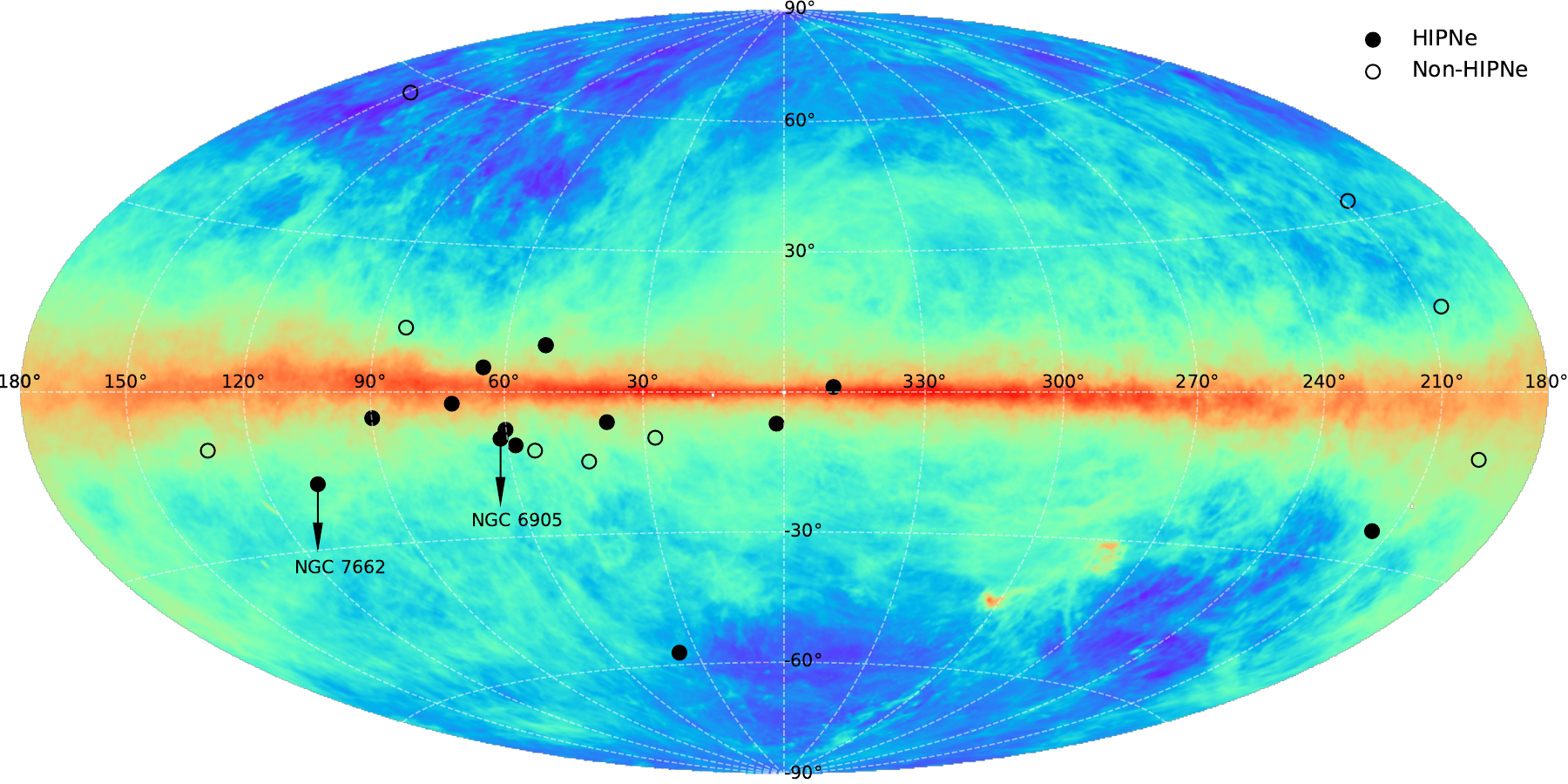}
\caption{Positions of the our PN sample and the known HIPNe 
retrieved from the literature in the Galactic coordinate system. 
The filled circles represent HIPNe, with newly detected sources in this observation labeled by their  names. The open circles represent PNe where
\ion{H}{1} is not detected in our observations.
The background is an all-sky \ion{H}{1} column density map integrated over the full velocity range of $-$600\,km~s$^{-1}$ $\leq$ $v_{\rm LSR}$ $\leq$ 600\,km~s$^{-1}$, adopted from
\citet{2016AA...594A.116H}.
 \label{fig:sample}}
\end{figure}

 While PNe near the Galactic disk typically host more massive progenitors that likely harbor significant \ion{H}{1} reservoirs, these regions suffer severe contamination from interstellar 21\,cm emission. To minimize this contamination, we selected 12 high-surface-brightness PNe at Galactic latitudes $|b| \geq 10\arcdeg$ (Figure~\ref{fig:sample}) for observation.
Table \ref{tab:para} lists the PNe in our sample, along with their coordinates and
other parameters. 
Column 7 presents the distances adopted in this work. Unless otherwise specified, the distances for the sources are 
primarily based on Gaia DR3 parallax measurements \citep{2021AA...656A.110C}.  
This step is carried out subsequent to verifying that the correct central star of the planetary nebula (CSPN) has been determined. \citet{2022Galax..10...32P} have demonstrated that up to 20{\%} of the GAIA-catalogued CSPN are misidentified. This is because a significant number of actual CSPN lie beyond the photometric limits of GAIA.
Columns 8 and 9 give the  angular diameters $\theta$ and expansion velocities $V_{\rm exp}$ of the PN.
We preferentially use the $\theta$ values provided by the HASH PN database \citep{2016JPhCS.728c2008P,2021MNRAS.503.2887B} if no other data is available.
Except for those specified, the $V_{\rm exp}$ values are taken from \cite{1984A&AS...58..273S}.

The observations were conducted from August 8 to November 27, 2021, using the 19-beam receiver of FAST in tracking 
mode. The integration time for each target varied from 40 to 50 minutes.
The 19-beam observation mode allows efficient sampling of spectra at on- and off-source positions simultaneously.
Each target PN was placed in the central beam (M01), which has a half-power beamwidth of 3$\arcmin$ at 1420 MHz and 
a pointing accuracy standard deviation of 7$\arcsec$.9. The surrounding beams (M02--07, see Papers I\&II for the beam
arrangement) sample the off-source positions about 6$\arcmin$ away from the PN in the east, south, west, and north 
directions. The peripheral beams (M08--19) are directed to the 
off-source positions about 12$\arcmin$ away from the center. The large separation between the on- and off-source 
positions ensures that no nebular emission is sampled in the off-position beams.

Spectra were recorded using the backend with 1,048,576 channels, resulting in a frequency resolution of 476.84\,Hz, 
which corresponds to a velocity resolution of 0.01 $\rm km~s^{-1}$ \citep{2020RAA....20...64J}.
For flux calibration, a high-intensity noise of 12\,K was periodically injected. 
 The antenna temperature  was converted to flux density using the telescope gain factor, as detailed in Section~2 of Paper~II. The {\it XX} and {\it YY} linear polarization spectra were individually calibrated and baseline-subtracted. 
 For the FAST, reflections off the dish surface and feed cabin generate standing waves that manifest as sinusoidal baselines, whereas broadband continuum emission yields polynomial-shaped baselines. We therefore selected either sinusoidal or polynomial baseline removal functions according to the specific baseline characteristics of each observation.
Because the baselines are relatively smooth,
their subtraction does not introduce significant uncertainty into the spectral measurements.
After calibration and baseline subtraction, the {\it XX} and {\it YY} spectra were averaged to obtain the final 
spectra. 

\section{Result} \label{sec:resu}

 Our goal is to detect \ion{H}{1} emission and absorption features in PNe, and we name objects displaying such spectral signatures as HIPNe.
If a feature at a right velocity
appears in the on-source spectrum but is invisible in all the off-source spectra, 
 we can consider it to potentially be of PN origin. However, this approach is only feasible when the interstellar emission varies smoothly across the beams. Otherwise, 
interferometric observations are more suitable to filter out the interstellar emission. Nevertheless, the 
single-dish observation provides an efficient way to search for HIPN candidates.

\begin{figure}[ht!]
\plotone{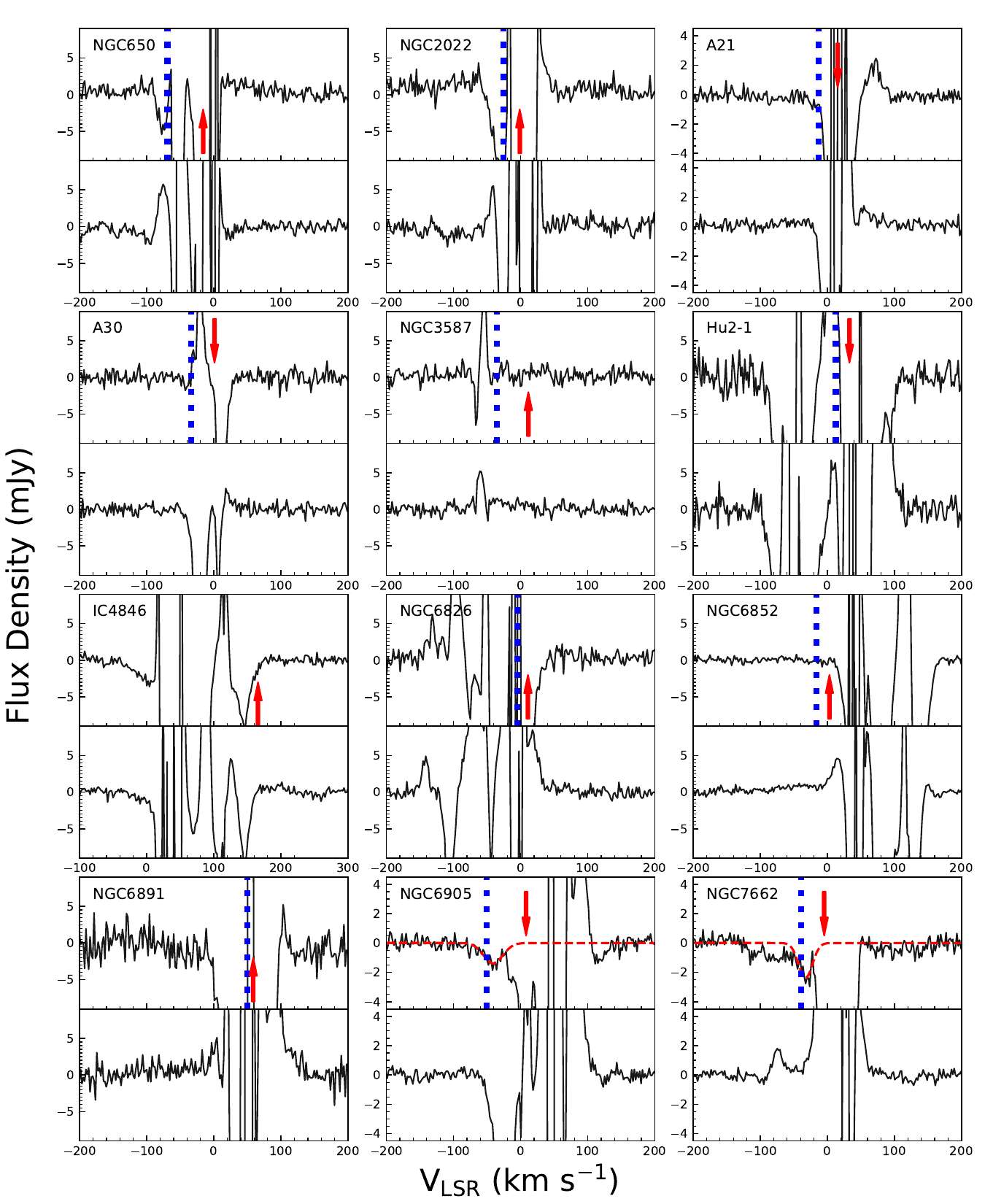}
\caption{\ion{H}{1} spectra of PNe. For each subfigure, the top and bottom panels show the ON$-$OFF and  OFF$-$OFF spectra, respectively.
The arrows mark the systemic velocity position the PNe.
The vertical dotted line indicates the velocities of the blue end of optical
emission lines ([\ion{N}{2}] or  H$\alpha$ lines),
 with the exception of IC 4846, where no such emission lines are found in the literature.
A fitting to the potential circumstellar \ion{H}{1} absorption is shown by the dashed curves.
\label{fig:spec}}
\end{figure}

Figure \ref{fig:spec} shows that the contamination from interstellar \ion{H}{1} emission is still significant for 
most of the PNe, largely hindering the detection of circumstellar  \ion{H}{1}.  Nevertheless, the systemic velocities
of NGC 3587 and NGC 6852 differ significantly from those of interstellar \ion{H}{1} foreground.  This allows us to derive
a tight upper limit of the flux density for the 21\,cm line
flux density ($S_{\rm line}$) in the two PNe,
based on the RMS of their spectra
($S_{\rm line} = \text{RMS}$).
For the other PNe, $S_{\rm line}$ is derived from the flux-density standard deviation of the M02--07 off-source spectra.

\subsection{H~I absorption}
 
Atomic hydrogen occupies the interface between the ionized and molecular regions of PNe, where the radiation field 
from the central star is not hard enough to ionize H but can significantly dissociate H$_2$.
The free-free emission arising from inner ionized nebulae provides the emission continuum at L band against which 
the 21\,cm \ion{H}{1} absorption could be measured blueward of the systemic velocity of the expanding ionized nebulae.

For spatially resolved PNe, the optical emission line profiles typically exhibit two split components. 
The blue-shifted component represents the part of the expanding nebula moving towards the observer, while the red-
shifted component represents the part moving away. The [\ion{N}{2}] lines at 6548 and 6584\,{\AA} can be used to
trace the velocity of the outermost ionized gas, which can be accelerated by the pressure of the ionized gas. Apart 
from [\ion{N}{2}] lines, the kinematic of the ionized gas's outer layers may also been traced by the 
velocity span of H$\alpha$ line \citep{2002A&A...384..620A}. 
We gathered the profiles of the  [\ion{N}{2}] lines
in the SPM Kinematic Catalogue of Galactic Planetary Nebulae\footnote{http://kincatpn.astrosen.unam.mx} 
\citep{2012IAUS..283...63L}  which covers many Galactic PNe of high surface brightness such as for this sample.
[\ion{N}{2}] lines  of NGC\,2022 and NGC\,6852 were  unavailable in the 
SPM catalogue, and thus we examined H$\alpha$ line instead. The catalogue does not include  IC\,4846.
As discussed in Paper II, the [\ion{N}{2}] lines trace the outermost ionized gas, which is spatially close to the neutral shell. Therefore, the circumstellar \ion{H}{1} absorption feature should be located 
on a position close to the blue end of the [\ion{N}{2}] and H$\alpha$ lines.

As shown in Figure \ref{fig:spec}, although interstellar \ion{H}{1} emission is strong, we tentatively detect 
circumstellar \ion{H}{1} absorption in NGC\,6905 and NGC\,7662. 
Taken the flux densities of the continuum emission at 21\,cm ($S_{\rm 21~cm}$) from the literature\footnote{Our observation setup was optimized for spectral line observations. The ON- and OFF-source spectra were acquired using different beams, introducing significant uncertainty in measuring the continuum emission. Additionally, the baseline may include both continuum emission and standing wave artifacts, 
prompting our reliance on literature-derived flux density values for subsequent analysis.} (as listed in 
column 4 of Table \ref{tab:cal}), we can derive 
the integrated optical depth using $\int \tau_v dv = -\int S_{\rm HI} dv/S_{\rm 21~cm}$, where $S_{\rm HI}$ is
the flux densities of the 21\,cm line subtracted by $S_{\rm 21~cm}$.
For non-detections, the upper limits of $|S_{\rm HI}|$
can be estimated from $S_{\rm line}$. Table \ref{tab:cal} presents $\int S_{\rm line} dv$ and the resultant $\int 
\tau_v dv$ upper limits in non-HIPNe. The circumstellar \ion{H}{1} absorption in HIPNe is measured
through a Gaussian profile fitting, and the results are presented in Table \ref{tab:ab}.
Based on the measured expansion velocities, angular radii, and distances, we determined the dynamic ages of the \ion{H}{1} shells are 2100 and 2400\,years in NGC 6905 and NGC 7662, respectively.

The \ion{H}{1} mass associated with the 21\,cm absorption can be obtained by the following formula:
\begin{equation}\label{m1}
M_{\rm ab} = 2.14\times10^{-6}\, \Gamma \left( \frac{T_{\rm ex}}{\rm K} \right)\left( \frac{D}{\rm kpc} \right)^2\left( \frac{\theta}{''} \right)^2\left( \frac{\int \tau_v dv}{\rm km\,s^{-1}} \right)\,{\rm M}_\sun,
\end{equation}
where $\Gamma$ is the geometric factor that accounts for atomic gases that are not in the line of sight, $T_{\rm ex}$ is the spin-excitation temperature, and $D$ is the distance.
For the calculations, we simply assume $\Gamma=1$ and $T_{\rm ex}=100$\,K 
(see Paper I for a discussion of the assumptions).
If the atomic gas is close to the ionization front,
$T_{\rm ex}$ approaches to 1000\,K, and $M_{\rm ab}$ would be underestimated by a factor of ten. 
If the atomic gas is distributed in the outer layers, similar in scale to H$_2$ (e.g. $\sim$2 times the size of  ionized regions), $\Gamma$ is 4, and 
$M_{\rm ab}$ would be underestimated by a factor of four. 
As a result, the  upper limits of $M_{\rm ab}$ range from 0.001 $\rm M_{\sun}$ to 16.2 $\rm M_{\sun}$. $M_{\rm ab}$ in HIPNe lies in a range of 0.01--0.14\,$\rm M_{\sun}$,  which encompasses the value of IC~4997 (Paper I).

\subsection{H~I emission}

Following the approach outlined in Papers I and II, we search
for \ion{H}{1}  emission by comparing the `ON-OFF' and `OFF-OFF' spectra. 
The off-position spectrum for primary comparison was obtained by averaging beam samples from positions M02--M07 (surrounding M01). Additionally, we generated two independent reference spectra by separately averaging the east--west and north--south beams.
However, we did not detect apparent \ion{H}{1} emission associated with the PNe. The mass upper limits of circumstellar \ion{H}{1} can be estimated from $S_{\rm line}$ using
the formula:
\begin{equation}\label{m2}
M_{\rm em} = 2.36\times10^{-4}\left( \frac{D}{\rm kpc} \right)^2\left( 
\frac{\int S_{\rm line} dv}{\rm mJy\,km\,s^{-1}} \right){\rm M}_\sun,
\end{equation}
as listed in Column 8 of Table~\ref{tab:cal}.
For comparison, we computed the mass of the ionized gas 
through 
\begin{equation}\label{m3}
M_{\rm i} = 5.34\times10^{-5}\left( \frac{D}{\rm kpc} \right)^{5/2}\left( \frac{\theta}{\arcsec} \right)^{3/2}\left( 
\frac{S_{\rm 6~cm}}{\rm mJy} \right){\rm M}_\sun,
\end{equation}
where $S_{\rm 6~cm}$ is the continuum emission at 6\,cm dominated
by free-free emission. The derived $M_{\rm i}$ values
 is listed in the column 10 of Table~\ref{tab:cal}, which
ranges from 0.005 to 1.9\,M$_{\sun}$.
Most of the PNe have $M_{\rm em}$ of less than 1\,M$_{\sun}$.
The tightest limit is that of NGC\,3578, in which $M_{\rm em}<0.004$\,M$_{\sun}$ and the mass ratio of atomic to ionized gas
is less than $0.01$.

In line with the cases of other PNe,
 \ion{H}{1}  emission was not detected  in the two HIPNe.
Table~\ref{tab:abei} presents the mass of the ionized gas and the upper
limits of the \ion{H}{1} emission gas in the HIPNe, 
as well as their morphological
classifications \footnote{The morphological classifications were adopted from the HASH database
\citep{2006MNRAS.373...79P}, 
which follows the classification scheme proposed by \cite{1995A&A...293..871C}. PNe are mainly categorized as Elliptical or oval (E), Round (R), Bipolar (B), Irregular (I), Asymmetric (A), or quasi-Stellar or point source (S) under the `ERBIAS' classifier. An additional `amprs' subclassifier is used to denote specific features, namely asymmetric enhancement (a), multiple shells (m), point-symmetry (p), ring structure or annulus (r), or resolved internal structure (s). Each PN is assigned one primary classification and may have multiple subclassifiers, which are listed in alphabetical order.}.
In principle, a comparison between $M_{\rm em}$ and $M_{\rm ab}$ could yield insights into the spatial extent of the \ion{H}{1}  shell. However, the  upper limits established for $M_{\rm em}$ exceed 
the $M_{\rm ab}$ values, precluding meaningful conclusions from this comparison.

\section{Discussion} \label{sec:disc}

The results allow us to investigate the `PN   mass missing problem'.
As pointed out by \citet{2012IAUS..283..362F},
the PN progenitor star has a typical mass of 2.5\,M$_\sun$. 
After the superwind phase, the mass of the residual white dwarf
is typically only 0.6\,M$_\sun$. 
Our results suggest that in general the mass of \ion{H}{1} gas are less or comparable with $M_i$. 
The upper limits of the total mass of ionized and atomic gas (based on \ion{H}{1} emission) are less than $\sim$2\,M$_\sun$ with  a median value of  0.9\,M$_\sun$. 
The mass ratios of atomic to ionized gas in NGC 6905 
and NGC 7662 are 0.3 and 0.02, respectively.
Therefore, the atomic gas cannot account for all the missing mass of PNe.

\begin{figure}[ht!]
\plotone{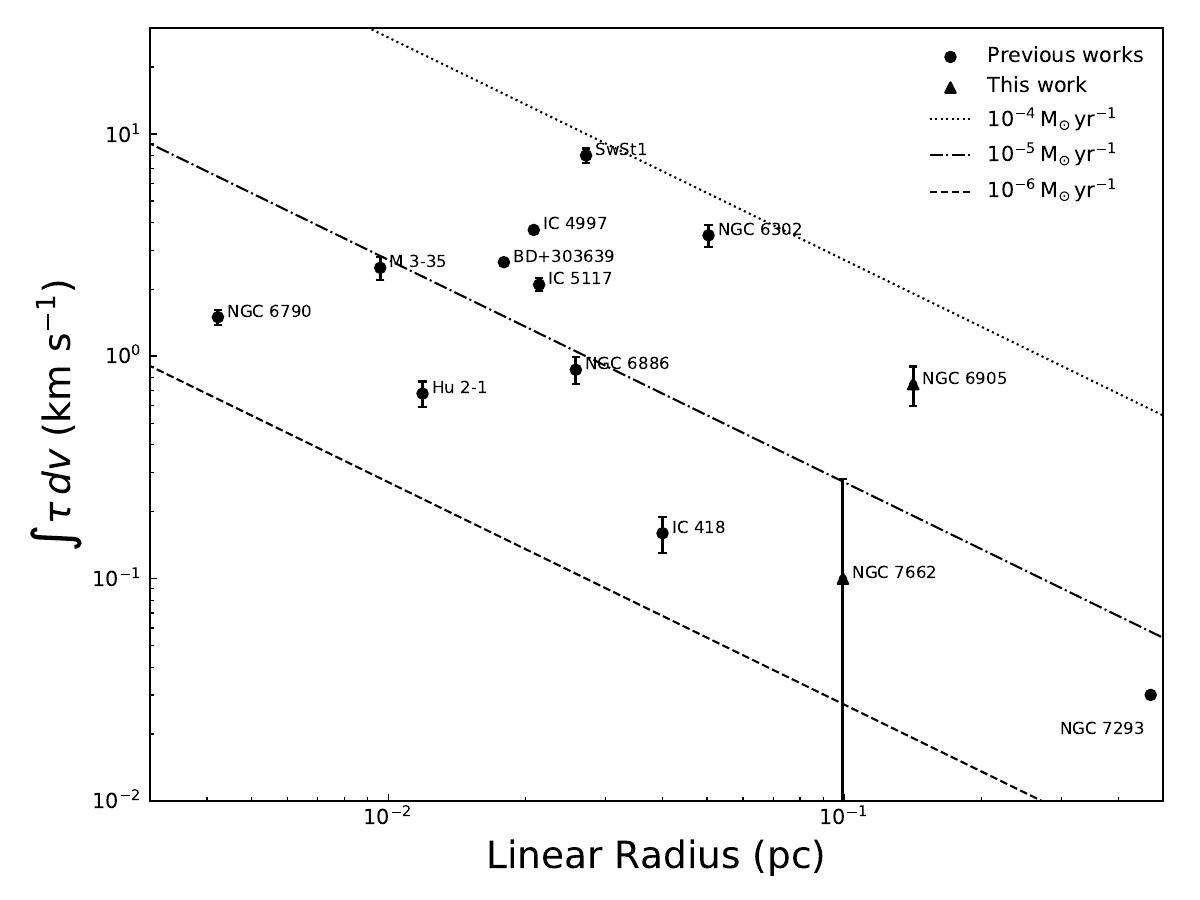}
\caption{Integrated \ion{H}{1} optical depth versus linear radius of the ionized regions.
The diagonal lines represent the results derived from theoretical 
calculations based on different mass-loss rates (as
shown in the upper right corner).
 \label{fig:depth}}
\end{figure}

It is conceivable that the optical depth of the neutral envelope 
increases with increasing mass loss rate ($\dot{M}$) and decreases
with nebular expansion. According to \citet{1990ApJ...351..515T}, they satisfy a relation of
\begin{equation}\label{m4}
\int \tau_v dv = 3.8 \times 10^6 \frac{\dot{M}}{T_{\rm ex} \mu v_n r_i} \left( 1 - \frac{r_i}{r_n} \right)\,{\rm 
km}\,{\rm s}^{-1},
\end{equation}
where $\mu$ is the mean molecular weight, $v_n$ is the expansion velocity of the neutral shell, and $r_i$ and $r_n$ 
are the radii of the ionized and neutral shells, respectively.
Assuming $r_n \gg r_i$, $v_n=10$\,km~s$^{-1}$, and $T_{\rm ex}=100$\,K, we calculate $\int \tau_v dv$
as functions of $r_i$ for $\dot{M} = 10^{-6}$, $10^{-5}$, and $10^{-4}$\,M$_\sun$ yr$^{-1}$, as indicated in 
Figure \ref{fig:depth}. Table \ref{tab:pab} lists all the PNe 
or which circumstellar \ion{H}{1} absorption has been detected in the existing literature, along with their 
respective characteristics. Here, $r_i$ is calculated by employing 
the updated distance values. As shown in Figure \ref{fig:depth}, the observational results
are consistent with the theoretical predictions, and indicate a mass loss rate of 10$^{-6}$--10$^{-4}$\,\,M$_\sun$ 
yr$^{-1}$ that is systematically higher than those obtained by 
\citet{1990ApJ...351..515T} and \citet{1995MNRAS.273..801G}. The high mass loss rates suggest that the \ion{H}{1} 
shell may be generated in the superwind epoch at the tip of the AGB. 

Our analysis of HIPNe (Tables~\ref{tab:abei} \& \ref{tab:pab}) reveals that 69{\%} (9 out of 13) exhibit bipolar morphologies (as classified in HASH).
This proportion is notably higher than 
the approximately
$\sim$20\% bipolar fraction observed in the general PN population \citep[e.g., HASH;][]{2011AJ....141..134S}. This correlation suggests a potential connection between bipolarity and enhanced \ion{H}{1} detectability.

The majority of prior detections of HIPNe 
have relied on single-dish observations, which lack 
sufficient  spatial resolution to resolve \ion{H}{1} structure. 
In contrast, interforemetric observations
of NGC\,6302 reveal that its \ion{H}{1} emission originates in a dust-shielded torus \citep{1985MNRAS.215..353R}, suggesting that such
circumstellar tori may protect neutral hydrogen from ionization. Importantly, all  HIPNe 
exhibit bright radio continuum emission in the NRAO VLA Sky Survey (NVSS) \citep{1998ApJS..117..361C}, 
a feature that provides a robust background for 
\ion{H}{1} absorption measurements and thereby enhances detection sensitivity.

\section{Conclusion} \label{sec:concl}

Using FAST, we conducted a \ion{H}{1} survey of 12  high surface brightness PNe at relatively high Galactic 
latitudes. The detection rate of HIPNe remains low due to either contamination from interstellar \ion{H}{1} emission or morphological selection bias in PN sample.
We report tentative detections of circumstellar \ion{H}{1} absorption in NGC\,6905 and NGC\,7662.
The mass of atomic gas in the two PNe ranges from  0.01 to 0.14\,$\rm M_\sun$, resulting in a mass ratio of 
atomic to ionized gas of 0.02--0.3. 

Nevertheless, our observations place tight upper limits for the  \ion{H}{1} mass in PNe, in particularly for those 
with  $V_{\rm LSR}$ substantially differing from the velocity of interstellar \ion{H}{1} lines. We show that the 
mass of atomic gas in PNe is unlikely to be higher than that of ionized mass, and thus 
the `PN mass missing problem' persists.

Furthermore, we confirm previous findings that the \ion{H}{1} absorption fades away with PN evolution. The high
mass loss rate implied by the observations suggests that  the \ion{H}{1} shell
may origin from a superwind epoch at the termination of AGB evolution.

The discovery of more HIPNe, strongly dominated by specific bipolar morphologies, is necessary to build a statistically significant sample to study the neutral
shell of PNe. Although interferometric observations has certain advantage of filtering out large-scale interstellar
\ion{H}{1} emission, sensitive single-dish observations such as FAST can offer a practical means of identifying
HIPN candidates. For that, we need to observe PNe with the line-of-sight velocity significantly 
different from the Galactic rotation  curve.

\section*{Acknowledgements}
We would like to express our sincere gratitude to the two anonymous reviewers for their insightful suggestions, which have significantly enhanced the quality of this paper.
The financial supports of this work are from 
the National Natural Science Foundation of China (NSFC, No.\,12473027 and 12333005),
the Guangdong Basic and Applied Basic Research Funding (No.\,2024A1515010798),
and the science research grants from the China Manned Space Project (NO. CMS-CSST-2021-A09, CMS-
CSST-2021-A10, etc). Y.Z. thanks the Xinjiang Tianchi Talent Program (2023).
X.H.L. acknowledges support from the Natural Science Foundation of Xinjiang Uygur Autonomous 
Region (No. 2024D01E37) and the National Science Foundation of China (12473025). Q.A.P. thanks 
the Hong Kong Research Grants Council for GRF research support under grants 17326116 and 17300417.
This work made use of the data from FAST (Five-hundred-meter Aperture Spherical radio Telescope). 
FAST is a Chinese national mega-science facility, operated by National Astronomical 
Observatories, Chinese Academy of Sciences.

%




\bibliography{sample631}{}
\bibliographystyle{aasjournal}



\newpage

\begin{deluxetable*}{lcccccccc}[htb]
\tabletypesize{\scriptsize}
\tablewidth{0pt}
\linespread{0.8}
\tablecaption{PN sample selected for the present work. \label{tab:para}}
\tablehead{\\
\colhead{Name} & \colhead{RA} & \colhead{DEC} & \colhead{$l$} & \colhead{$b$} & \colhead{$V_{\rm LSR}$} & \colhead{$D^d$} & \colhead{$\theta$} & \colhead{$V_{\rm exp}$}\\
\colhead{} & \colhead{(h m s)} & \colhead{($\arcdeg$ $\arcmin$ $\arcsec$)} & \colhead{($\arcdeg$)} & \colhead{($\arcdeg$)} & \colhead{($\rm km~s^{-1}$)} & \colhead{kpc} & \colhead{$\arcsec$} & \colhead{($\rm km~s^{-1}$)}\\
\colhead{(1)} & \colhead{(2)} & \colhead{(3)} & \colhead{(4)} & \colhead{(5)} & \colhead{(6)} & \colhead{(7)} & \colhead{(8)} & \colhead{(9)}
}
\startdata
    NGC\,650-1 & 01h42m19.66s & +51d34m31.6s & 130.9 & $-$10.5 & $-16.1 \pm 1.2^a$  &  $0.93 \pm 0.26^e$   & $84.00^f$  &  38.5\\
    NGC\,2022 & 05h42m06.19s & +09d05m10.9s & 196.7 & $-$10.9 & $-1.3 \pm 2.2^a$  &  $2.27^{+0.29}_{-0.23}$  &  $23.22^{g}$  &  27.5\\
    Abell\,21   & 07h29m02.88s & +13d14m30.0s & 205.1 & +14.2 & $15.4  \pm 5.2^a$  &   $0.59^{+0.03}_{-0.02}$  & $375.00^f$  &  $32.0^{h}$\\
    Abell\,30   & 08h46m54.40s & +17d52m32.6s & 208.6 & +33.3 & $0.9^b$  &   $2.22^{+0.16}_{-0.14}$  & $63.50^f$  &  40.0\\
    NGC\,3587 & 11h14m47.71s & +55d01m08.5s & 148.5 & +57.0 & $11.5  \pm 3.1^a$  &   $0.81^{+0.03}_{-0.03}$  & $104.00^f$  &  30.0\\
    Hu\,2-1 & 18h49m47.57s & +20d50m39.5s & 51.5 & +9.7 & $33.0  \pm 3.1^a$  &   $2.38^{+0.48}_{-0.34}$  &  $2.06^{g}$  &  10.0\\
    IC\,4846 & 19h16m28.22s & $-$09d02m36.8s & 27.6 & $-$9.6 & $165.7 \pm 3.0^a$  &  $7.13 \pm 2.01^e$   &  $2.93^{g}$  &  $13.1^{i}$\\
    NGC\,6826 & 19h44m48.15s & +50d31m30.2s & 83.6 & +12.8 & $10.9 \pm 0.6^a$  &  $1.30^{+0.07}_{-0.06}$   &  $21.96^{g}$  &  8.0\\
    NGC\,6852 & 20h00m39.21s & +01d43m40.9s & 42.6 & $-$14.5 & $-3.1^c$  &   $2.56^{+1.14}_{-0.60}$  &  $1.67^{g}$  &  $43.1^{h}$\\
    NGC\,6891 & 20h15m08.84s & +12d42m15.6s & 54.2 & $-$12.1 & $58.6 \pm 1.0^a$  &   $2.56^{+0.21}_{-0.18}$  &  $9.91^{g}$  &  10.0\\
    NGC\,6905 & 20h22m22.99s & +20d06m16.3s & 61.5 & $-$9.6 & $8.1   \pm 1.7^a$  &   $2.70^{+0.24}_{-0.20}$  &  $21.65^f$  &  43.5\\
    NGC\,7662 & 23h25m53.60s & +42d32m06.0s & 106.6 & $-$17.6 & $-4.7  \pm 0.7^a$  &   $1.75^{+0.10}_{-0.09}$  &  $23.42^{g}$  &  26.5\\
\enddata
\tablecomments{$^a$ \cite{1983ApJS...52..399S}; $^b$ \cite{1989IAUS..131..183J}; $^c$ \cite{1988ApJ...334..862M}; $^d$ \cite{2021AA...656A.110C}; $^e$ \cite{2016MNRAS.455.1459F}; $^f$ HASH; $^{g}$ \cite{2021MNRAS.503.2887B}; $^{h}$ \cite{2021RAA....21..151A}; $^{i}$ \cite{1992AA...260..314B}.}
\end{deluxetable*}

\begin{deluxetable*}{lccccccccc}[htb]
\tabletypesize{\scriptsize}
\linespread{0.8}
\tablecaption{Spectroscopic measurements and   upper mass limits of  \ion{H}{1} in the non-HIPNe. \label{tab:cal}}
\tablehead{\\
\colhead{Name} &  \colhead{rms} &  \colhead{Devi} &  \colhead{$S^b_{\rm 21\,cm}$} &  \colhead{$S_{\rm 6\,cm}$} & \colhead{$\int S_{\rm line} dv$} & \colhead{$\int\tau_v dv$} &  \colhead{$M_{\rm em}$} &  \colhead{$M_{\rm ab}$} &  \colhead{$M_{\rm i}$}\\
\colhead{} & \colhead{(mJy)} & \colhead{(mJy)} & \colhead{(mJy)} & \colhead{(mJy)} & \colhead{($\rm mJy~km~s^{-1}$)} & \colhead{($\rm km~s^{-1}$)} & \colhead{(M$_\sun$)} & \colhead{(M$_\sun$)} & \colhead{(M$_\sun$)}\\
\colhead{(1)} & \colhead{(2)} & \colhead{(3)} & \colhead{(4)} & \colhead{(5)} & \colhead{(6)} & \colhead{(7)} & \colhead{(8)} & \colhead{(9)} & \colhead{(10)}
}
\startdata
    NGC\,650-1 & 0.74 &  34.44  & $141.1\pm4.9$  & $117^d$ & 1414.5  & 10.0  & $\textless$0.3  & $\textless$3.3  & 0.4 \\
    NGC\,2022 & 0.93 &  51.51  & $93.2\pm3.8$  & $80^d$  & 1511.1 & 16.2  & $\textless$1.8 &  $\textless$2.4 & 0.4 \\
    Abell\,21   & 0.33 &  15.40  & $\textgreater85^c$  & $327^e$  & 525.7 & $\textless$6.2  & $\textless$0.04 & $\textless$16.2  & 1.9 \\
    Abell\,30   & 0.77 &  5.50  & $\textless5$  & \nodata  & 234.7 & $\textgreater$46.9  & \nodata & \nodata  & \nodata \\
    NGC\,3587 & 0.74 &  0.42  & $115.9\pm6.8$  & $88^d$  & 23.68 & 0.2  & $\textless$0.004 & $\textless$0.08  & 0.3 \\
    Hu\,2-1$^a$ & 1.58 &  121.53  & $43.0\pm1.4$  & $110^f$  & 1296.4 & 30.1  & $\textless$1.7 & $\textless$0.04  & 0.01 \\
    IC\,4846 & 0.48 &  1.2  & $40.7\pm1.6$  & $43^e$  & 16.7 & 0.4  & $\textless$0.2 & $\textless$0.01  & 0.2 \\
    NGC\,6826 & 0.77 &  4.46  & $415.0\pm16.0$  & $373^d$  & 38.1 & 0.1  & $\textless$0.02 & $\textless$0.004  & 0.2 \\
    NGC\,6852 & 0.34 &  1.04  & $13.5\pm1.5$  & $20^e$  & 15.63 & 1.2  & $\textless$0.02 &  $\textless$0.001 & 0.005 \\
    NGC\,6891 & 1.88 &  61.77  & $111.3\pm3.4$  & $109^d$  & 658.9 &  5.9 & $\textless$1.0 &  $\textless$0.2 & 0.2 \\
\enddata
\tablecomments{$^a$
\cite{1995MNRAS.273..801G} reported tentative \ion{H}{1} detection in this bipolar PN. However, in the present work,
the detection of \ion{H}{1} is significantly obstructed by the substantial contamination of interstellar \ion{H}{1} emission.
$^b$ \cite{1998ApJS..117..361C}; $^c$ \cite{1999ApJS..123..219C}; $^d$ \cite{1991ApJS...75....1B}; $^e$ \cite{1979AAS...36..227M}; $^f$ \cite{1990AAS...84..229A}.}
\end{deluxetable*}

\begin{deluxetable*}{lcccccccc}[htb]
\tabletypesize{\scriptsize}
\tablewidth{0pt}
\linespread{0.8}
\tablecaption{Measurements of the \ion{H}{1} absorption  and \ion{H}{1} masses in the HIPNe.
 \label{tab:ab}}
\tablehead{\\
\colhead{Name} &  \colhead{$S^a_{\rm 21\,cm}$} &  \colhead{$\int S_{\rm HI} dv$} &  \colhead{$V_{\rm HI}$} &  \colhead{FWHM} &  \colhead{$S_{\rm p}$} & \colhead{$\int\tau_v dv$} & \colhead{$M_{\rm ab}$} & \colhead{$V_{\rm exp}$}\\
\colhead{} & \colhead{(mJy)} & \colhead{($\rm mJy~km~s^{-1}$)} & \colhead{($\rm km~s^{-1}$)} & \colhead{($\rm km~s^{-1}$)} & \colhead{($\rm mJy$)} & \colhead{($\rm km~s^{-1}$)} & \colhead{(M$_\sun$)} & \colhead{($\rm km~s^{-1}$)}\\
\colhead{(1)} & \colhead{(2)} & \colhead{(3)} & \colhead{(4)} & \colhead{(5)} & \colhead{(6)} & \colhead{(7)} & \colhead{(8)} & \colhead{(9)}}
\startdata
    NGC\,6905 & $67.3\pm2.7$ & $-50.40\pm5.59$   & $-40.75\pm1.67$  & $33.91\pm4.74$ & $-1.40\pm0.12$ & $0.75\pm0.15$ & 0.14 & 65.80 \\
    NGC\,7662 & $646.0\pm19.0$ & $-61.78\pm9.34$  & $-32.52\pm1.51$  & $24.39\pm4.74$ & $-2.38\pm0.25$ & $0.10\pm0.18$ & 0.01 & 40.01 \\
\enddata
\end{deluxetable*}

\begin{deluxetable*}{lccccccc}[htb]
\tabletypesize{\scriptsize}
\linespread{0.8}
\tablecaption{
Ionized gas masses and $M_{\rm em}$ upper limits
in the HIPNe.
\label{tab:abei}}
\tablehead{\\
\colhead{Name} &\colhead{rms} &  \colhead{Devi} &  \colhead{$S_{\rm 6\,cm}$} & \colhead{$\int S_{\rm line} dv$} &  \colhead{$M_{\rm em}$} &  \colhead{$M_{\rm i}$} & \colhead{Morph.}\\
\colhead{}& \colhead{(mJy)} & \colhead{(mJy)} & \colhead{(mJy)} & \colhead{($\rm mJy~km~s^{-1}$)} & \colhead{(M$_\sun$)} & \colhead{(M$_\sun$)} & \\
\colhead{(1)} & \colhead{(2)} & \colhead{(3)} & \colhead{(4)} & \colhead{(5)} & \colhead{(6)} & \colhead{(7)}  & \colhead{(8)}
}
\startdata
    NGC\,6905 & 0.34 &  2.31  & $63^a$  & 107.2 & $\textless$0.18  & 0.51 & Bmps \\
    NGC\,7662 & 0.46 &  12.06  & $605^a$  & 340.9  & $\textless$0.25  & 0.60 & Emrs \\
\enddata
\tablecomments{$^a$ \cite{1991ApJS...75....1B}. 
}
\end{deluxetable*}

\begin{deluxetable*}{lcccccccc}[htb]
\tabletypesize{\scriptsize}
\tablewidth{0pt}
\linespread{0.8}
\tablecaption{Catalog of HIPNe from the literature.
\label{tab:pab}}
\tablehead{\\
\colhead{Name} &\colhead{$l$}  & \colhead{$b$} &  \colhead{Distance} &  \colhead{$\theta^l$} &  \colhead{$r_i$} &  \colhead{$\int\tau_v dv$} & \colhead{Morph.}\\
\colhead{}& \colhead{$\arcdeg$} & \colhead{$\arcdeg$} & \colhead{(kpc)} & \colhead{($\arcsec$)} & \colhead{(pc)} & \colhead{($\rm km~s^{-1}$)} & \\
\colhead{(1)} & \colhead{(2)} & \colhead{(3)} & \colhead{(4)} & \colhead{(5)} & \colhead{(6)} & \colhead{(7)} & \colhead{(8)}}
\startdata
    NGC\,6302 & 349.51 & +1.06 & 1.17$^a$   & 17.77  & 0.05  & $3.5\pm0.4$$^{n}$  & Bamps \\
    NGC\,6790 & 037.89 & -6.31 & 0.8$^b$   & 2.18  & 0.04  & $1.5\pm0.12$$^{o}$  & Bps \\
    NGC\,6886 & 60.14 & -7.73 & 2.6$^c$   & 4.09  & 0.03  & $0.87\pm0.12$$^{o}$  & Bmps \\
    IC\,418 & 215.21 & -24.28 & 1.26$^d$   & 13.08  & 0.04  & $0.16\pm0.03$$^{o}$  & Emrs \\
    IC\,5117 & 89.87 & -5.13 & 5.02$^e$   & 1.76  & 0.02  & $2.1\pm0.14$$^{o}$  & Bmrs \\
    SwSt1 & 1.59 &  -6.72 & 2$^f$   & 5.6$^{m}$  & 0.03  & $8\pm0.6$$^{p}$  & S \\
    Hu\,2-1 & 51.48 & +9.68 & 2.38$^g$   & 2.06  & 0.01  & $0.68\pm0.09$$^{p}$  & Bmps \\
    M\,3-35 & 71.63 & -2.36 & 2.4$^{h}$   & 1.65  & 0.01  & $2.5\pm0.3$$^{p}$  & Bamps \\
    NGC\,7293 & 36.16 & -57.12 & 0.2$^{i}$   & 970$^{m}$  & 0.47  & 0.03$^{q}$  & Bmprs \\
    IC\,4997 & 58.33 & -10.98 & 5$^j$   & 1.72  & 0.02  & $3.7\pm0.06$$^{j}$  & Bmps \\
    BD+30$\arcdeg$3639 & 64.79 & +5.02 & 1.5$^k$   & 4.93  & 0.02  & $2.65\pm0.09$$^{k}$ & Eamrs \\
\enddata
\tablecomments{$^a$ \cite{2008AJ....135.2074G}; $^b$ \cite{1986AA...157..191G}; $^c$ \cite{2005AA...432..139P}; $^d$ \cite{2009AA...507.1517M}; $^e$ \cite{2015MNRAS.447.1673V}; $^f$ \cite{2001MNRAS.328..527D}; $^g$ \cite{2022AA...658A..17S}; $^{h}$ \cite{1989AAS...78..301W}; $^{i}$ \cite{2018AA...616A...9L}; $^j$ \cite{2022ApJ...933....4O}; $^k$ \cite{2023ApJ...952..166O}; $^l$ \cite{2021MNRAS.503.2887B}; $^{m}$ HASH; $^{n}$ \cite{1982Natur.299..323R}; $^{o}$ \cite{1990ApJ...351..515T}; $^{p}$ \cite{1995MNRAS.273..801G}; $^{q}$ \cite{2002ApJ...574..179R}. }
\end{deluxetable*}

\end{document}